\documentclass[%
reprint,
superscriptaddress,
amsmath,amssymb,
]{revtex4-1}

\usepackage{graphicx}
\usepackage{dcolumn}
\usepackage{bm}
\usepackage{hyperref}

\newcommand{\calH}{\mathcal{H}}

\newcommand{\calL}{\mathcal{L}}

\newcommand{\calD}{\mathcal{D}}

\usepackage{braket}
\usepackage{color}

\begin{document}

\preprint{APS/123-QED}

\title{
Dielectric breakdown of strongly correlated insulators in one dimension: \\ Universal formula from non-Hermitian sine-Gordon theory
}

\author{Kazuaki Takasan}
\email{kazuaki.takasan@phys.s.u-tokyo.ac.jp}
\affiliation{Department of Physics, University of California, Berkeley, California 94720, USA}
\affiliation{Materials Sciences Division, Lawrence Berkeley National Laboratory, Berkeley, California 94720, USA}
\affiliation{Department of Physics, University of Tokyo, 7-3-1 Hongo, Bunkyo-ku, Tokyo 113-0033, Japan}

\author{Masaya Nakagawa}
\affiliation{Department of Physics, University of Tokyo, 7-3-1 Hongo, Bunkyo-ku, Tokyo 113-0033, Japan}

\author{Norio Kawakami}
\affiliation{Department of Physics, Kyoto University, Kyoto 606-8502, Japan}
\affiliation{Fundamental Quantum Science Program, TRIP Headquarters, RIKEN, Wako 351-0198, Japan}
\affiliation{Department of Physics, Ritsumeikan University, Kusatsu, Shiga 525-8577, Japan}
\affiliation{Department of Materials Engineering Science, Osaka University, Toyonaka, Osaka 560-8531, Japan}

\date{\today}

\begin{abstract}
Application of a strong electric field to insulators induces a finite current. This phenomenon is called dielectric breakdown and is known as a fundamental nonequilibrium and nonlinear transport phenomenon in solids. Here, we study the dielectric breakdown of generic strongly correlated insulators in one dimension. Combining bosonization techniques with a quantum tunneling theory, we develop an effective field-theoretical description of dielectric breakdown using a non-Hermitian sine-Gordon theory. Then, we derive an analytic formula for the threshold field, which is a many-body generalization of the Landau-Zener formula. Importantly, we point out that the threshold field contains a previously overlooked factor originating from the charges of elementary excitations, which should be significant when a system has fractionalized excitations. We apply our results to integrable lattice models and confirm that our formula is valid in a broad range including the weak coupling regime, indicating its wide potential applicability. Our results unveil universal aspects of nonlinear and nonequilibrium transport phenomena in various strongly correlated insulators.
\end{abstract}

\maketitle

\section{Introduction}

It is an important subject in condensed matter physics and statistical physics to understand nonlinear transport phenomena that cannot be described by linear response theory~\cite{NonlinearTransport_Book, Bonitz2000}. 
One of the typical nonlinear transport phenomena is \textit{dielectric breakdown}. Applying a sufficiently strong electric field makes insulators conductive and induces a finite current. For band insulators, it is known as Zener breakdown and is well-understood~\cite{Zener1934}. In contrast, dielectric breakdown in strongly correlated insulators is a relatively new topic~\cite{Fukui1998, Taguchi2000} and has been gathering great attention~\cite{Oka2003, Oka2005LZ, Oka2010, Eckstein2010, Meisner2010, Oka2012, Aron2012, Lenarcic2012}. 
The main focus in this field has been on fermionic Mott insulators. For example, theoretical studies on the Hubbard model have revealed that the dielectric breakdown can be described by a many-body generalization of the Landau-Zener tunneling~\cite{Oka2003, Oka2005LZ, Oka2010, Oka2012}. From the experimental side, starting from pioneering work on SrCuO$_2$~\cite{Taguchi2000}, dielectric breakdown phenomena have been explored in various materials, e.g., vanadium oxides~\cite{Kalcheim2020} and organic materials~\cite{Yamakawa2017, Takamura2023}. More recently, the crossover between DC and AC driving~\cite{Li2022, Shinjo2024} and the effect of nearest-neighbor interactions~\cite{Hansen2025} have been investigated, and the connection to quantum electrodynamics~\cite{Oka2005LZ} has been revisited~\cite{Queisser2025}. For a recent review of nonequilibrium phenomena in Mott insulators under an electric field, see Ref.~\cite{MurakamiRMP2025}.

In contrast to the intensive studies on fermionic Mott insulators, dielectric breakdown phenomena in more generic strongly correlated insulators, such as charge-density-wave (CDW) insulators, bosonic Mott insulators, and Kondo insulators, have not been fully understood yet, and thus the universal properties common in these systems remain elusive~\cite{Sachdev2002, Simon2011, Queisser2012, Kolovsky2016, Zhu2018}. This is because it is hard to treat nonequilibrium dynamics and many-body problems simultaneously and thus theoretical approaches are mostly limited to numerical calculations.

\begin{figure}[t]
\includegraphics[width=8.8cm]{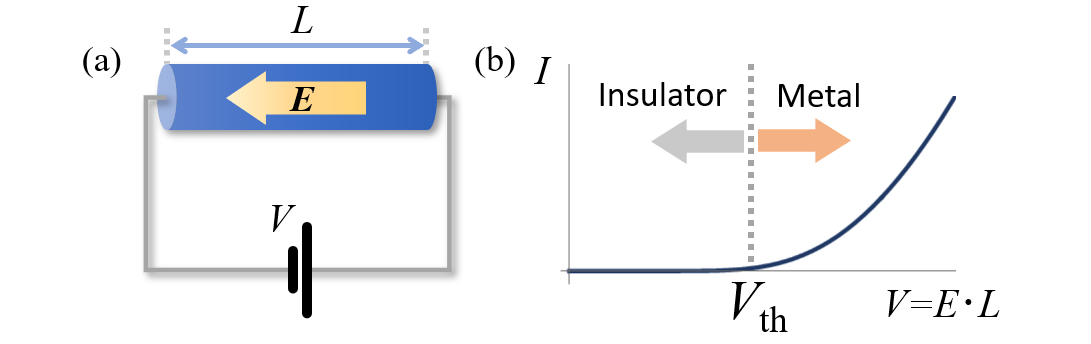}
\caption{
(a) Our setup of a strongly correlated one-dimensional insulator under a DC electric field $\bm E$.
(b) A typical current-voltage relation ($I$-$V$ characteristic) of dielectric breakdown phenomena. At the threshold voltage $V_\mathrm{th}$, the nonequilibrium insulator-metal transition occurs. 
}
\label{Fig:setup}
\end{figure}

In this paper, we present an analytic approach for resolving this issue in one dimension. Our approach is based on the low-energy effective field theory (EFT), which enables us to treat a broad range of one-dimensional (1D) strongly correlated insulators, not limited to the fermionic Mott insulators.
The setup is schematically shown in Fig.~\ref{Fig:setup}~(a). Utilizing a quantum tunneling theory~\cite{Dykhne1962, Davis1976} and bosonization techniques~\cite{Giamarchi_book, Tsvelik_book, Gogolin_book}, we derive an EFT that describes the dielectric breakdown phenomena in strongly correlated insulators. This EFT is a \textit{non-Hermitian sine-Gordon theory} that includes an imaginary vector potential. On the basis of this theory, we derive an analytic formula for the threshold electric field, which is the most important quantity characterizing the dielectric breakdown. This formula can be interpreted as a many-body generalization of the famous Landau-Zener formula~\cite{Landau1932, Zener1932, Zener1934}. 
Also, this formula is a generalization of the one derived in Ref.~\cite{Oka2012} for the fermionic Hubbard model~\footnote{Our result is a generalization of the results in Ref.~\cite{Oka2012} for the following two reasons: One is that our result is applicable to a wide range of one-dimensional insulators while the previous result is limited to fermionic Mott insulators. The other is that our result contains the new input $e/e^*$ which has been overlooked because $e/e^*$ is unity in the fermionic Hubbard model.}. Remarkably, our formula suggests that fractionalization in elementary excitations has significant effects on the threshold field, which cannot be captured by the original Landau-Zener formula. 
Furthermore, to clarify the validity of our EFT, we apply the results to several solvable lattice models. We find that our formula shows good agreement with microscopic calculations across a wide range including the weak coupling regime, indicating its wide potential applicability. 
Note that our results can be applied to generic models, including non-integrable models, because the derivation is based solely on the low-energy EFT and does not refer to any information about specific lattice models.

The field-theoretical description developed here allows us to discuss the universal aspects of dielectric breakdown in one-dimensional many-body quantum systems in a unified fashion. In addition, our formula should be useful for understanding experiments related to dielectric breakdown in strongly correlated insulators beyond simple fermionic Mott insulators.

\section{Model and Methods}
To investigate 1D strongly correlated insulators [Fig.~\ref{Fig:setup}~(a)], we start from a generic lattice model subject to a DC electric field~\cite{FNH}, 
\begin{align}
H(t) =-\sum_{i \alpha} \left( e^{iaA(t)} c^\dagger_{i \alpha} c_{i+1 \alpha} + \mathrm{h.c.}  \right) + V_\mathrm{int}. \label{eq:lattice_model}
\end{align}
Here, the electric field $E = V/L$ ($V$: applied voltage, $L$: system size) is included in the time-dependent vector potential $A(t)=-Ft$, with $F=eE$ and $a$ denoting a lattice constant. $c^\dagger_{i \alpha}$ ($c_{i \alpha}$) denotes a creation (annihilation) operator of a particle at the $i$-th site with some internal degrees of freedom, e.g., spins or orbitals, denoted by $\alpha$. We assume periodic boundary conditions in this study. For later convenience, we introduce a time-dependent basis $\ket{\phi_n (t)}$ that satisfies $H(t) \ket{\phi_n (t)} = E_n (t) \ket{\phi_n (t)}$ ($n=0, 1, 2, \cdots$), where $\ket{\phi_{0(n)} (t)}$ corresponds to the ground ($n$-th excited) state~\cite{FNA}.  We assume that the interaction term $V_\mathrm{int}$ makes the system fully gapped, i.e., $\bar{\Delta} (t) \equiv E_1(t) - E_0(t) > 0$ for all $t \in \mathbb{R}$, and insulating at zero temperature.

The real-time dynamics in this model is described by the Schr\"odinger equation $i \frac{d}{dt} \ket{\psi (t)} = H (t) \ket{\psi (t)}$ with the initial state $\ket{\psi (0)} = \ket{\phi_0 (0)}$. In this time evolution, the strong electric field induces excitations from the ground state $\ket{\phi_0 (t)}$ to the excited states $\ket{\phi_n (t)}$ ($n > 0$) which can carry finite currents. This transition occurs repeatedly over the time period $T = 2 \pi  / (F L)$, and thus the survival probability of the ground state $\mathcal{P}_0=|\braket{\phi_0(mT)|\psi(mT)}|^2$ is approximately given as $(1-p)^m$, where $p$ is the transition probability from the ground state to the first excited state, since this process is the most dominant~\cite{FN2}. With increasing field strength $F$, the probability $p$ starts to grow abruptly at the threshold field $F_\mathrm{th}$, similar to the current $I$, as shown in Fig.~\ref{Fig:setup}~(b), and the system is easily driven to the current-carrying states. This behavior corresponds to the breakdown phenomenon. For convenience, we rewrite the probability as $p = \exp ( - \pi F_\mathrm{th} / F )$. This formula provides the definition of the threshold field $F_\mathrm{th}$, which has already been adopted in previous studies on fermionic Mott insulators~\cite{Oka2003, Oka2005LZ, Oka2010, Oka2012}. Note that this threshold behavior is nothing but the signature appearing in the $I$-$V$ characteristics in Fig.~\ref{Fig:setup}~(b) and the current $I$ is given as $I \sim F \exp (-\pi F_\mathrm{th}/F)$ as discussed in Ref.~\cite{Oka2012}.

To calculate the probability $p$, we use the Dykhne-Davis-Pechukas (DDP) formula~\cite{Dykhne1962, Davis1976}, which provides the non-adiabatic tunneling probability approximately for generic two-level quantum systems~\cite{Suominen1991, Vasilev2004, Kitamura2020, Fukushima2020, Takayoshi2020}. It has already been applied to breakdown phenomena in several quantum many-body systems~\cite{Oka2010, Oka2012, Uchino2012, Tripathi2016, Lankhorst2018} in good agreement with other numerical approaches~\cite{Oka2010, Oka2012}. Based on the DDP formula, the tunneling probability to the first excited state is given as $p = \exp \left( -2 \mathrm{Im} \int^{t_c}_0 \bar{\Delta}(t) dt \right)$, where $t_c$ is a critical time defined by $\bar{\Delta} (t_c) = 0$, which takes a complex value since $\bar{\Delta}(t) > 0$ for any real $t$. Changing the variable from the time $t$ to the vector potential $A= - F t$, we obtain $F_\mathrm{th}= (2/\pi) \mathrm{Im} \int^{A_c}_0 \Delta (A) d A$, where $\Delta (A) \equiv \bar{\Delta}(t=A/F)$ and $A_c = - F t_c$. Since $t_c$ is complex, $A_c$ also becomes complex, i.e., $A_c = i h_c + c$ ($h_c$ and $c$ are real). In a sufficiently large system with an energy gap, the dependence of the energy on the real part of a vector potential is negligible~\cite{Oka2010, Watanabe2018} and thus the threshold field is rewritten as
\begin{align}
F_\mathrm{th} &= \frac{2}{\pi} \int^{h_c}_0 \mathrm{Re} \Delta (ih) dh. \label{eq:Fth}
\end{align}
Below, we evaluate this formula~(\ref{eq:Fth}) for generic 1D systems (\ref{eq:lattice_model}). 
For this purpose, we need to analyze the model with an imaginary vector potential $A=ih$, reflecting the extension of time to complex values. After plugging $A=ih$ into Eq.~(\ref{eq:lattice_model}) [i.e., $e^{\pm iaA} \rightarrow e^{\mp ah}$], the hopping becomes asymmetric, thereby making the model \textit{non-Hermitian}~\cite{FNB}.
A similar approach for quantum tunneling problems with non-Hermitian models was used in Refs.~\cite{Oka2010, Oka2012, Uchino2012, Tripathi2016, Lankhorst2018, Suominen1991, Vasilev2004, Fukushima2020, Kitamura2020, Takayoshi2020}. 
This type of asymmetric hopping, which also appears in the Hatano-Nelson model~\cite{Hatano1996}, has been extensively studied in the context of non-Hermitian physics~\cite{AshidaNonHermitianReview2020}.

\section{The effective field theory}
\label{sec:EFT}
To obtain the threshold field, we need to calculate the energy gap of the non-Hermitian model. However, except for integrable lattice models, it is difficult to analyze generic models. To overcome this difficulty, we derive an EFT that describes the low-energy behavior. Although the full nonequilibrium dynamics may not be described with the EFT, the threshold field (\ref{eq:Fth}) is determined from the low-energy part and thus allows a field-theoretical description, as shown below. In 1D systems, bosonization techniques provide a systematic framework for the derivation of effective field theories~\cite{Haldane1981, Giamarchi_book, Tsvelik_book, Gogolin_book}.

Following the standard procedure of the bosonization, as outlined in Appendix~\ref{sec:bosonization}, we obtain an action $S[\phi]$ of the EFT as
\begin{align}
S[\phi] &=\frac{1}{2 \pi K}  \int d\tau dx \left \{ \frac{1}{v} (\partial_\tau \phi)^2 + v  (\partial_x \phi)^2 \right \} \nonumber \\
&\qquad\quad+ \frac{h}{\pi} \int d\tau dx (\partial_\tau \phi)+  g \int d\tau dx \cos ( \beta \phi), \label{eq:sine_Gordon_S_with_h}
\end{align}
which is a non-Hermitian sine-Gordon theory. Here, $\phi$ is a bosonic field that describes the density fluctuations related to the density of electrons as $\rho=-(1/\pi)\partial_x\phi$, $v$ is the velocity of collective charge excitations, and $K$ is the Luttinger parameter. This sine-Gordon model [Eq.~(\ref{eq:sine_Gordon_S_with_h})] consists of two parts. One is the free-boson part (the first term) with an imaginary vector potential (the second term), which represents the hopping term in Eq.~(\ref{eq:lattice_model}) with $A=ih$ and describes the metallic behavior. The other part is the cosine term (the third term), which denotes the interaction term in Eq.~(\ref{eq:lattice_model}). The cosine term pins the bosonic field to its potential minimum and opens an energy gap in the insulator. Therefore, the sine-Gordon model is regarded as an effective model for describing metal-insulator (gapless-gapped) transitions and is applied to various problems in 1D quantum many-body physics~\cite{Giamarchi_book, Tsvelik_book, Gogolin_book}. The parameters $v, K, g$ are generally renormalized due to the interaction effect, whereas $\beta$ is fixed in each lattice model~(\ref{eq:lattice_model}). The partition function at zero temperature is defined as $Z = \int D\phi e^{-S[\phi]}$ using the above action~\cite{FNE}.

\section{Formula for the threshold field}

From Eq.~(\ref{eq:Fth}), we can see that there are two tasks for deriving the threshold field: (i) One is to find the critical value $h_c$, and (ii) the other is to find $\mathrm{Re} \Delta (ih)$, which is the change in the real part of the energy gap as a function of the imaginary vector potential. After completing them, we can obtain the analytic form of the threshold field following from the formula (\ref{eq:Fth}).

First, we calculate the critical value $h_c$. 
To this end, we perform a \textit{space-(imaginary-)time transposition}, i.e., $(\tilde{x}, \tilde{\tau})= (v \tau, x/v)$, on the action (\ref{eq:sine_Gordon_S_with_h}). Introducing a new field as $\tilde{\phi} (\tilde{x}, \tilde{\tau}) \equiv \phi(v \tilde{\tau}, \tilde{x}/v)$, we obtain the action for $\tilde{\phi} (\tilde{x}, \tilde{\tau})$. With redefining $\phi (x, \tau) = \tilde{\phi} (\tilde{x}, \tilde{\tau})$, this action takes the same form as the original action with the following replacement 
\begin{align}
\frac{h}{\pi}\int d\tau dx (\partial_\tau \phi) \rightarrow \frac{vh}{\pi}\int dx d\tau  (\partial_{x} \phi ). \label{eq:replace}
\end{align}
Remarkably, this model is regarded as a model of static insulators with doping represented by the chemical potential $\mu = vh$ because the bosonized form of the chemical potential term is given as $-\mu \int dx d\tau \rho (x, \tau) = \frac{\mu}{\pi} \int dx d\tau \partial_{x} \phi (x, \tau)$~\cite{Giamarchi_book}. Since the space-time transposition does not change the partition function $Z$, 
the models before and after the transformation exhibit common critical properties and show the same metal-insulator transitions with changing $h$ or $\mu$ respectively. To obtain the critical value, we rewrite the term $-(vh/\pi) \int dx \partial_x \phi$ as $- (2vh/\beta) Q_\mathrm{top}$ where $Q_\mathrm{top} \equiv \beta/(2\pi) \int dx \partial_x \phi$ is the topological charge of the excitations in the sine-Gordon theory. Because the elementary excitation in the sine-Gordon theory is given by a (anti-)soliton with mass $M$ and topological charge $+1$ ($-1$)~\cite{Samaj_book}, the excitation becomes gapless when $2vh/\beta$ reaches $M$. Thus, the critical point is given as $h_c = \beta M / (2v)$. Using the energy gap $\Delta_0 = 2M$ which is the energy cost to create a pair of soliton and anti-soliton, we obtain
\begin{align}
h_c = \frac{e}{e^*} \cdot \frac{\Delta_0}{2v}, \label{eq:h_c}
\end{align}
where $ - e^* \equiv - 2e/\beta$ is the elementary physical (not topological) charge of a soliton excitation~\cite{FN5, FNF}. 
Note that this result provides a generic proof of a conjecture proposed by Nakamura and Hatano~\cite{Nakamura2006, FN4}. 

\begin{figure}[t]
\includegraphics[width=8.5cm]{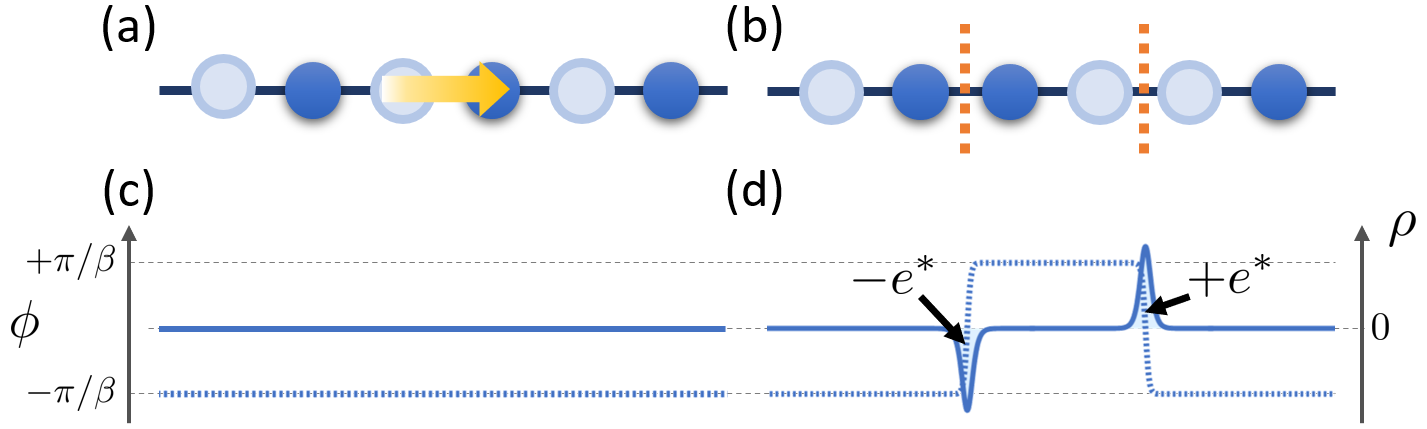}
\caption{
(a, b) Schematic picture of pair creation of a soliton and an anti-soliton. The filled (empty) circle denotes a particle (hole). The figure (a) ((b)) is before (after) the pair creation. The yellow arrow and the orange broken lines represent the electric field and the position of the domain walls respectively. (c, d) The field configuration (broken line) and the soliton density (solid line). (c) and (d) correspond to (a) and (b) respectively. First, the field is pinned to one of the potential minima and there is no excitation. After the pair creation, two kinks are generated in the field. The shaded area corresponds to the charge of the (anti-)soliton $e^*$ ($-e^*$).
}
\label{Fig:soliton}
\end{figure}

Before proceeding to the next task, we comment on this mapping from the asymmetric hopping model to the doped static insulator. This suggests that the dielectric breakdown has a dual theoretical description with an insulator-to-metal transition induced by doping. This is quite nontrivial because this mapping relates seemingly different kinds of metal-insulator transitions, one of which occurs in nonequilibrium, and the other occurs in equilibrium. Furthermore, it is surprising that this mapping reduces the non-Hermitian problem to the Hermitian one. Due to the time-reversal symmetry of the non-Hermitian Hamiltonian~\cite{FNJ}, the partition function $Z$ is real and the existence of the corresponding Hermitian model is allowed.

Next, we work on the second task (ii). 
As we mentioned, elementary excitations in the sine-Gordon theory are a soliton and an anti-soliton and the dielectric breakdown in the sine-Gordon model corresponds to a pair creation of a soliton and an anti-soliton (schematically shown in Fig.~\ref{Fig:soliton}). Therefore, the energy gap between the ground state and the first-excited state is written as $\Delta(\theta) = E_s(\theta) + E_{\bar{s}}(\theta)$ where $E_{s(\bar{s})}(\theta)$ is the energy dispersion of a soliton  (anti-soliton) with its rapidity $\theta$ and already known in integrable field theories~\cite{Samaj_book}. This energy dispersion $E_{s(\bar{s})}(p)$ is a relativistic one given as the energy $E_{s(\bar{s})}(\theta) = M \cosh \theta$ and the momentum $p(\theta) = (M/v) \sinh \theta$. 
Therefore, we reach $\Delta(ih) = 2 M \cosh \theta(h)$ and then we need the rapidity of the soliton under an imaginary vector potential $A=ih$. For this purpose, we consider a wave function of a single soliton $\varphi_{\theta}(x) \propto \exp [- i p(\theta) x]$. Because the vector potential can be rewritten as the twisted boundary condition with an imaginary twist angle, the effect of the imaginary vector potential is represented as $\varphi_{\theta} (L) = \exp[- e^* L h / e] \varphi_\theta (0)$. Therefore, we obtain $p(\theta) = i e^* h /e$, and then $\theta = \sinh^{-1} [iv e^* h/(eM)]$. Using the energy gap $\Delta_0 = 2M$, we arrive at 
\begin{align}
\Delta (ih) = \Delta_0 \sqrt{1-\left(\frac{e^*2vh}{ e\Delta_0}\right)^2} = \Delta_0 \sqrt{1-\left(\frac{h}{h_c}\right)^2}, \label{eq:Delta(ih)}
\end{align}
reflecting the relativistic dispersion of solitons in the sine-Gordon theory.

\section{Universal formula of the threshold field}
Finally, we combine the above results to obtain the threshold field $F_\mathrm{th}$. We can perform the integral in Eq.~(\ref{eq:Fth}) with Eq.~(\ref{eq:Delta(ih)}) as $F_\mathrm{th} = (2 \Delta_0 h_c/\pi) \int^{1}_0  \sqrt{1-x^2} dx =\Delta_0 h_c/2$. Substituting Eq.~(\ref{eq:h_c}), we obtain the threshold field as
\begin{align}
F_\mathrm{th} = \frac{e}{e^*}\cdot  \frac{(\Delta_0/2)^2}{v}.\label{eq:MBLZ}
\end{align}
This is one of our main results in this study. 
Eq.~(\ref{eq:MBLZ}) gives a universal formula applicable to any 1D strongly correlated insulator described by the sine-Gordon model. This formula also extends the field-theoretical description of the equilibrium universality class of 1D insulators to the dielectric breakdown, which is a typical nonequilibrium phenomenon. 

Furthermore, this formula can be regarded as a many-body generalization of that for band insulators obtained by Zener~\cite{Zener1934}. Indeed, our formula has a similar form to the threshold field of the Landau-Zener formula $F^{\mathrm{LZ}}_\mathrm{th}=(\Delta_\mathrm{LZ}/2)^2/v_0$ where the non-adiabatic transition probability is given by $p= \exp ( - \pi F^{\mathrm{LZ}}_\mathrm{th} / F )$ for a time-dependent two-level system $H(t)=\begin{pmatrix} v_0 t & \Delta_\mathrm{LZ}/2 \\ \Delta_\mathrm{LZ}/2 & -v_0 t \end{pmatrix}$~\cite{Landau1932, Zener1932}. However, our formula contains remarkable features beyond the Landau-Zener formula, reflecting many-body effects. One is that the quantities $\Delta_0$ and $v$, which are the many-body energy gap and the velocity of the collective excitations, fully include the many-body effects~\cite{FN9}. The other is the factor $e/e^*$, which corresponds to the inverse of the physical charge of elementary excitations. This indicates that insulators hosting fractional excitations (i.e., $e^* < e$) have a larger threshold field. For example, this effect should appear in CDW insulators hosting domain-wall-type excitations with fractional charges as shown in Fig.~\ref{Fig:soliton}~\cite{Su1979, Su1980, Rice1982, FNC}.

\begin{figure}[t]
\includegraphics[width=8.7cm]{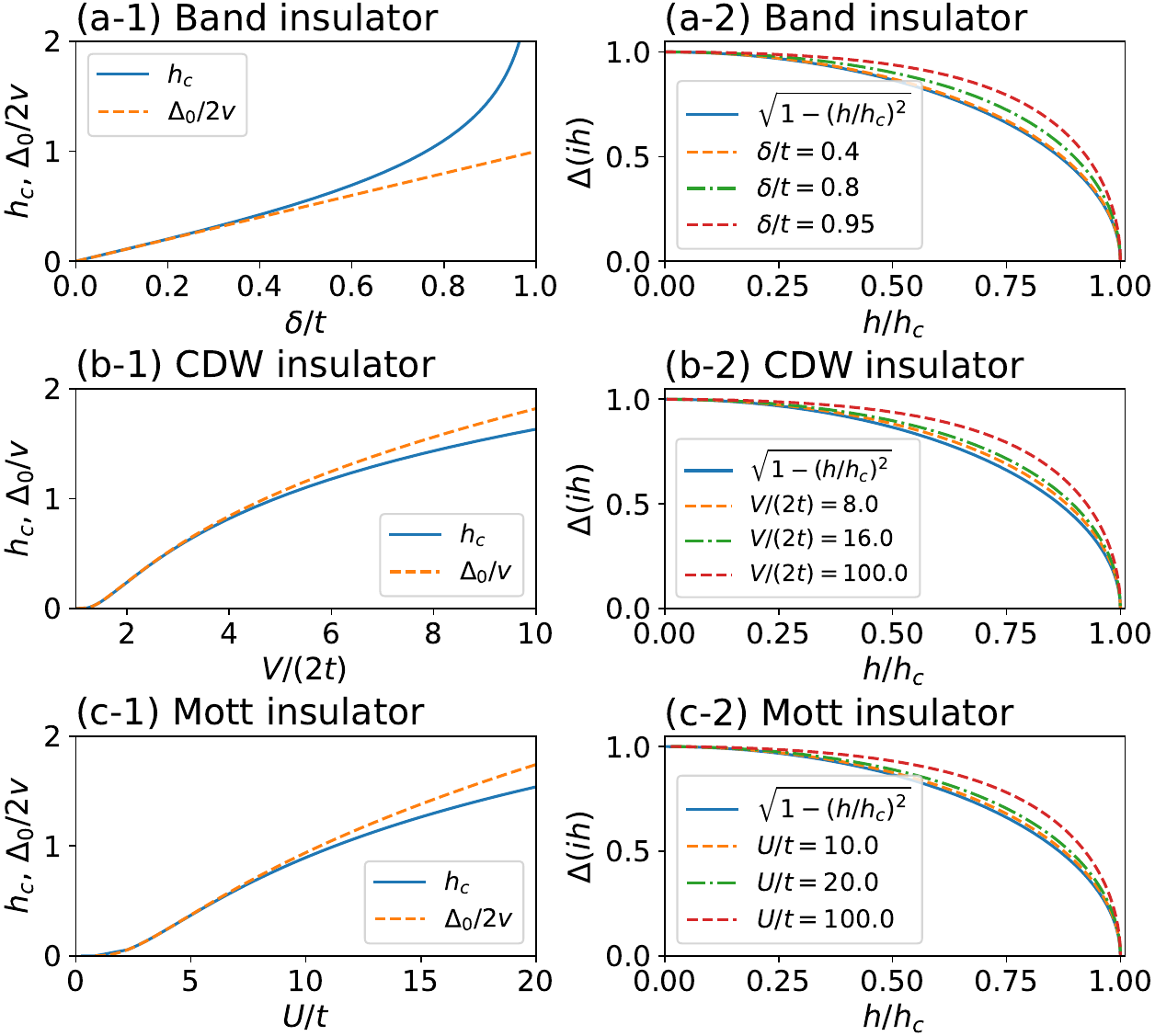}
\caption{Comparison between the predictions [Eqs.~(\ref{eq:h_c}) and (\ref{eq:Delta(ih)})] based on the effective field theory and the exact results calculated from integrable lattice models. (a) is for a band insulator (Su-Schrieffer-Heeger model), (b) is for a CDW insulator (spinless fermions with nearest neighbor interaction), and (c) is for a Mott insulator (Fermi-Hubbard model). The critical value $h_c$ together with the right hand side of Eq.~(\ref{eq:h_c}) and the change of energy gap $\Delta (ih)/\Delta_0$ together with $\sqrt{1-(h/h_c)^2}$ are shown in (X-1) and (X-2) respectively (X = a, b, c).}
\label{Fig:lattice}
\end{figure}

\section{Applications to integrable lattice models} 
Our derivation is based on the low-energy EFT and the resulting relations in Eqs.~(\ref{eq:h_c}) and (\ref{eq:Delta(ih)}) are closely related to the emergent Lorentz invariance. Therefore, it is unclear whether our formula is valid in the original lattice model where the Lorentz invariance is violated. To clarify to what extent the formula is valid, we apply it to integrable lattice models. We calculate the critical value $h_c$, the energy gap $\Delta_0$, the velocity $v$, and the energy gap under an imaginary vector potential $\Delta(ih)$, and then check Eq.~(\ref{eq:h_c}) and Eq.~(\ref{eq:Delta(ih)}) respectively. The details of the calculation are presented in Appendix~\ref{sec:integrable}. We study the Su-Schrieffer-Heeger model, $H_\mathrm{SSH} = - \sum_i(t+(-1)^i\delta)\{c^\dagger_ic_{i+1}+\mathrm{h.c.}\}$, spinless fermions with a nearest neighbor interaction, $H_\mathrm{SLF}=-t\sum_{i}(c^\dagger_{i+1}c_i+\mathrm{h.c.})+V\sum_i(c^\dagger_{i+1}c_{i+1}-1/2)(c^\dagger_{i}c_{i}-1/2)$, and the Fermi-Hubbard model, $H_\mathrm{Hub}=-t\sum_{i\sigma}(c^\dagger_{i\sigma}c_{i+i \sigma}+\mathrm{h.c.})+U\sum_ic^\dagger_{i\uparrow}c_{i\uparrow}c^\dagger_{i\downarrow}c_{i \downarrow}$ where $c_{i}$ ($c_{i \sigma}$) denotes an annihilation operator of spinless fermions (fermions with spin $\sigma = \uparrow, \downarrow$) at the $i$-th site. These models are integrable lattice models and their low-energy properties are described by the sine-Gordon theory. Each model exemplifies a band insulator, a CDW insulator, and a fermionic Mott insulator respectively. The results are shown in Fig.~\ref{Fig:lattice} and it is clearly seen that both Eq.~(\ref{eq:h_c}) and Eq.~(\ref{eq:Delta(ih)}) are valid over a broad range including the weak coupling regime. Note that the critical field $h_c$ of the CDW insulator contains the effect of the fractionalized charge $e/e^*=2$ as shown in Fig.~\ref{Fig:lattice}~(b-1).
In the strong coupling regime, the lattice effect becomes significant and the threshold field deviates from our field-theoretical prediction~\cite{FNG}. This implies that the naive application of the Landau-Zener formula to strong coupling regimes gives incorrect predictions. 

\section{Discussion}
While most previous studies have focused on integrable lattice models (e.g., the Fermi-Hubbard model~\cite{Fukui1998, Oka2010, Oka2012, Uchino2012}) to obtain an analytic result, our approach does not rely on the integrability of lattice systems, which can be easily broken in realistic situations.
Indeed, there are various insulators that cannot be described by an integrable lattice model. One example is a Kondo insulator, which is a strongly correlated insulator realized in heavy fermion systems~\cite{Coleman_book}. This insulator is known to be described by a Kondo lattice model, which is considered to be non-integrable but can be described using bosonization techniques~\cite{Fujimoto1994, Fujimoto1997, Tsunetsugu1997, Nakagawa2017}. Another example is a bosonic Mott insulator, which is realized with ultracold bosonic atoms in optical lattices~\cite{Greiner2002}.
Such bosons are described by a Bose-Hubbard model, which is also non-integrable but can be described by the sine-Gordon model via phenomenological bosonization~\cite{Haldane1981, Giamarchi_book, FND}. We leave the numerical comparison of our analytic results in these non-integrable models for future studies.

\section{Conclusion and Outlook}
We have investigated the dielectric breakdown in generic strongly correlated insulators. For this purpose, we have constructed an EFT based on a theory of quantum tunneling and bosonization techniques and derived the universal formula of threshold fields which is widely applicable to strongly correlated insulators in one dimension. Our important finding is that the formula contains a charge of elementary excitations, which suggests that the dielectric breakdown may provide useful information about fractionalized quasiparticles in interacting systems. Furthermore, to clarify the range where our theory works, we have applied our formula to integrable lattice models and found good agreement with their exact solutions. This supports the validity of our field-theoretical description. We hope that this study provides a step toward understanding universal aspects of nonequilibrium phenomena and nonlinear transport in quantum many-body systems.

\begin{acknowledgments}
We are thankful to Yuto Ashida, Yohei Fuji, Shunsuke Furukawa, Shunsuke Furuya, Hosho Katsura, Takashi Oka, Shintaro Takayoshi, and Masahito Ueda for valuable discussions. This work was supported by a Grant-in-Aid for Scientific Research on Innovative Areas “Topological Materials Science” (KAKENHI Grant No. JP15H05855), JSPS KAKENHI (Grants Nos. JP16J05078, JP18H01140, JP19H01838, JP20K14383, JP22K20350, JP23K17664, JP24K16989, and JP25K17312), and JST PRESTO (Grant Nos. JPMJPR2256 and JPMJPR2596). M.N. was supported by RIKEN Special Postdoctoral Researcher Program. K.T. thanks JSPS for support from Research Fellowship for Young Scientists and Overseas Research Fellowship.
\end{acknowledgments}

\appendix

\section{Derivation of the effective action}
\label{sec:bosonization}

In this section, we outline the derivation of the effective action (\ref{eq:sine_Gordon_S_with_h}). 

\subsection{Bosonization formulae and sine-Gordon model}
There is a powerful theoretical tool called \textit{bosonization} for one-dimensional quantum systems. This enables us to systematically derive low-energy effective field theories and has led to many successful results in one-dimensional quantum many-body problems~\cite{Giamarchi_book}. Here, we use bosonization techniques to derive an effective field theory from the lattice model [Eq.~\eqref{eq:lattice_model}].

The most important relationship in bosonization is \textit{the bosonization formula}, which allows us to represent any operators in one-dimensional models with bosonic field operators corresponding to the low-energy excitations. We write down this formula below, following the notation in Giamarchi's textbook~\cite{Giamarchi_book}. The operator of a spinless fermion $c_i$ in a one-dimensional chain with the lattice constant $a$ is written with bosonic fields $\phi(x)$ and $\theta(x)$ as \begin{align}
\frac{c_i}{\sqrt{a}} = c (x) &= \psi_R (x) + \psi_L (x),\\
\psi_r (x) &= e^{i r k_F x} \tilde{\psi}_r (x) \nonumber\\
&= \frac{U_r}{\sqrt{2 \pi \alpha}} e^{i r k_F x} e^{-i(r \phi(x) - \theta (x))}, \label{eq:BF_SLF}
\end{align}
where $r = R(+1), L(-1)$ represents a right-mover and a left-mover, $U_r$ is a Klein factor expressed in terms of Majorana fermions, $\{ U_r, U_{r^\prime} \} = 2 \delta_{r r^\prime}$, and $\alpha$ denotes a short-range cutoff. 
For fermions with spins $s$ ($= \uparrow, \downarrow$), a fermion operator $c_{is}$ in a one-dimensional chain with the lattice constant $a$ reads
\begin{align}
\frac{c_{is}}{\sqrt{a}} = c_s (x) &= \psi_{Rs} (x) + \psi_{Ls} (x),\\
\psi_{rs} (x) &= e^{i r k_F x} \tilde{\psi}_{rs} (x)\nonumber\\
&= \frac{U_{r s}}{\sqrt{2 \pi \alpha}} e^{i r k_F x} e^{-i(r \phi_s(x) - \theta_s (x))}.\label{eq:BF_SFF}
\end{align}
For convenience, we define the charge sector ($\rho$) and the spin sector ($\sigma$) as
\begin{align}
\phi_{\rho} (x) = \frac{\phi_\uparrow (x) + \phi_\downarrow (x)}{\sqrt{2}}, \phi_{\sigma} (x) = \frac{\phi_\uparrow (x) - \phi_\downarrow (x)}{\sqrt{2}}, \label{eq:phi_chargespin}\\
\theta_{\rho} (x) = \frac{\theta_\uparrow (x) + \theta_\downarrow (x)}{\sqrt{2}}, \theta_{\sigma} (x) = \frac{\theta_\uparrow (x) - \theta_\downarrow (x)}{\sqrt{2}}.
\end{align}
It is also possible to obtain a bosonization formula for bosonic systems using phenomenological bosonization~\cite{Giamarchi_book}. The density operator $\rho(x)$ and an annihilation operator of a boson $ b_i $ are represented with bosonic fields $\phi (x)$ and $\theta (x)$ as
\begin{align}
\rho(x) &= \rho_0 - \frac{1}{\pi} \nabla \phi(x) + \rho_0 \sum_{p \neq 0} e^{i 2p (\pi \rho_0 x - \phi (x))}, \\
\frac{b_{i}}{\sqrt{a}} &= \rho^{1/2}_0 \sum_{p} e^{-i 2p (\pi \rho_0 x - \phi (x))} e^{i \theta(x)}. \label{eq:BF_Boson}
\end{align}

Applying the above formulae to the lattice model (\ref{eq:lattice_model}) with $A(t) = 0$ and taking a continuum limit ($a \to 0$), we obtain the low-energy effective theory of the lattice model. As mentioned in Sec.~\ref{sec:EFT}, the hopping part in Eq.~(\ref{eq:lattice_model}) gives a gapless free boson theory and the interaction term in Eq.~(\ref{eq:lattice_model}) is typically recast to a cosine term. While it is possible that multiple cosine terms appear depending on the details of the interaction term, we consider a single cosine term for simplicity. Then, we obtain the following Hamiltonian, called \textit{sine-Gordon model},
\begin{align}
H = \frac{v}{2 \pi}\int d x \left\{ K (\pi \Pi)^2 + \frac{1}{K} ( \nabla \phi )^2 \right\} +  g \int d x \cos ( \beta \phi).
\end{align}
Here we have defined the conjugate momentum $\Pi = \nabla \theta(x) / \pi$ which satisfies $[\phi(x), \Pi(y)]=i \delta(x-y)$.  $K, v, g$ and $\beta$ are model parameters. $K$ is called the Luttinger parameter and $v$ corresponds to the velocity of bosonic excitations. $g$ represents the strength of the interaction inducing the energy gap. Due to the interaction effect represented by the cosine term, $K$, $v$ and $g$ take renormalized values, whereas $\beta$ is fixed in each lattice model~(\ref{eq:lattice_model}).

\subsection{Sine-Gordon model with a real/imaginary vector potential}

Next, we introduce the vector potential to the sine-Gordon model. The vector potential provides a shift of the momentum $\Pi$ as
\begin{align}
\Pi \rightarrow \Pi - \frac{A}{\pi}.
\end{align}
Indeed, using the bosonization formula [Eq.~(\ref{eq:BF_SLF}), (\ref{eq:BF_SFF}), or (\ref{eq:BF_Boson})], this substitution corresponds to the insertion of the $U(1)$ gauge flux $\Phi = A L$ to the system whose size is $L$, i.e., the twisted boundary condition with the twist angle $\Phi$~\cite{Giamarchi_book}. Note that when $\phi$ represents the charge sector $\phi_\rho (x)$, $A$ is replaced by $\sqrt{2} A$ due to the factor $1/\sqrt{2}$ appearing in the definition of the charge sector [Eq.~(\ref{eq:phi_chargespin})]. Introducing the vector potential in this way, the Hamiltonian becomes 
\begin{align}
H = \frac{v}{2 \pi}\int \!\! d x \left\{ K (\pi \Pi - A)^2 \!+\! \frac{1}{K} ( \nabla \phi )^2 \right\} +  g \!\! \int \!\! dx \cos ( \beta \phi),
\end{align}
where the integration is performed from 0 to $L$ and we consider the thermodynamic limit $L \to \infty$. 

We use the space-time transposition to derive Eq.~(\ref{eq:h_c}). For this purpose, we move to the Lagrangian formulation via the Legendre transformation. Using the Hamiltonian density $\calH$ defined from $H = \int d x \calH$, the partition function at zero temperature is defined as $Z = \int \calD \phi \calD \Pi e^{-\tilde{S}[\phi, \Pi]
}$ with the action $\tilde{S}[\phi, \Pi] = \int dx d\tau \tilde{\calL}[\phi, \Pi]$, where the integral along the $\tau$-direction is performed from 0 to $\infty$, and the Lagrangian density $\tilde{\calL} = - i \Pi \partial_\tau \phi + \calH$. The Lagrangian density is rewritten as 
\begin{align}
\tilde{\calL}[\phi, \Pi] &= - i \Pi \partial_\tau \phi + \frac{v K}{2\pi} (\pi \Pi - A)^2 \nonumber \\
&\quad + \frac{v}{2 \pi K} (\partial_x  \phi)^2 + g \cos (\beta \phi) \nonumber \\
&= \frac{vK\pi}{2} \left \{ \Pi^2 - 2 \left( \frac{i}{vK\pi} (\partial_\tau \phi) + \frac{A}{\pi} \right) \Pi \right\} \nonumber \\
&\quad+ \frac{vK}{2\pi} A^2 + \frac{v}{2 \pi K} (\partial_x  \phi)^2 + g \cos (\beta \phi).
\end{align}
Performing the Gaussian integral for the momentum $\Pi$, we obtain the new action $S$ satisfying $Z = \int \calD \phi e^{-S [\phi]}$ with $S = \int dx d\tau \calL [\phi]
$ and the new Lagrangian density $\calL$ is written as
\begin{align}
\calL[\phi] &= - \frac{vK\pi}{2} \left( \frac{i}{vK\pi} (\partial_\tau \phi) + \frac{A}{\pi} \right)^2 + \frac{v}{2 \pi K} (\partial_x  \phi)^2  \nonumber\\
&\quad+ \frac{vK}{2\pi} A^2+ g \cos (\beta \phi)\nonumber \\
&=\frac{1}{2 \pi K} \left\{ \frac{1}{v} (\partial_\tau \phi )^2 + v (\partial_x \phi)^2 \right\} \nonumber \\
&\quad- i \frac{A}{\pi} (\partial_\tau \phi) + g \cos (\beta \phi).
\end{align}
Using this Lagrangian density, we obtain the action
\begin{align}
S[\phi] &= \frac{1}{2 \pi K}\int d\tau dx \left \{ \frac{1}{v} (\partial_\tau \phi)^2 + v  (\partial_x \phi)^2 \right \}\nonumber \\
&\quad- i\frac{A}{\pi} \int d\tau dx (\partial_\tau \phi)+  g \int d\tau dx \cos ( \beta \phi).
\end{align}
Substituting the imaginary vector potential $A= ih$ into this action, we reach the action (\ref{eq:sine_Gordon_S_with_h}).

\section{Details of the analysis of integrable lattice models}
\label{sec:integrable}

To show the validity of our formula [Eqs.~(\ref{eq:h_c}) and (\ref{eq:Delta(ih)})] based on the effective field theory, we confirm these equations in the specific integrable lattice models. We treat three models below: 1. Su-Schrieffer-Heeger model, 2. Spinless fermions with nearest-neighbor interaction (equivalent to an XXZ spin chain), 3. Fermi-Hubbard chain.

\subsection{Su-Schrieffer-Heeger model}

The Su-Schrieffer-Heeger model is defined as 
\begin{align}
H_\mathrm{SSH} = \sum_i (-t + (-1)^i \delta) \left\{ c^\dagger_i c_{i+1} + \mathrm{h.c.} \right\}. \label{eq:SSH_lattice}
\end{align}
Using bosonization techniques~\cite{Giamarchi_book}, we can obtain the bosonized form of the above Hamiltonian
\begin{align}
H_\mathrm{SSH}  &= \frac{v}{2 \pi}\int d x \left\{ (\pi \Pi)^2 + ( \nabla \phi )^2 \right\} +  g \int d x \cos ( 2\phi), \label{eq:SSH_SG}
\end{align}
with $v=2at$ and $g=2\delta/\alpha$. Examining the cosine term of the model~(\ref{eq:SSH_SG}), we find that the parameter $\beta$ appearing in Eq.~\eqref{eq:sine_Gordon_S_with_h} is equal to 2 and thus $-e^* = -e$, which corresponds to a usual electron excitation. 

To confirm Eqs.~(\ref{eq:h_c}) and (\ref{eq:Delta(ih)}), we calculate the energy gap $\Delta_0$, the critical value $h_c$, the velocity $v$, and the change of the energy gap $\Delta(ih)$. Since this model is a model of free fermions, we can easily diagonalize the Hamiltonian and obtain the above quantities. 

To obtain the energy gap $\Delta_0$, we rewrite the Hamiltonian (\ref{eq:SSH_lattice}) in the momentum representation as $H_\mathrm{SSH} (k) = (-2 t \cos ak ) \sigma_x + (2 \delta \sin ak ) \sigma_y$ and diagonalize it. Then, we obtain the energy eigenvalues as $E_\pm(k, \delta)= \pm t \sqrt{2 (1+\bar{\delta}^2) + 2 (1 - \bar{\delta}^2) \cos 2 ak}$ where we define $\bar{\delta}=\delta/t$. Using these eigenvalues, the energy gap is calculated as $\Delta_0 = E_+(\pm \pi/(2a), \delta) - E_-(\pm \pi/(2a), \delta)= 4\delta$. Thus, the quantity $(e/e^*)\Delta_0/(2v)$ is calculated as 
\begin{align}
\frac{e}{e^*}\cdot\frac{\Delta_0}{2v} = \frac{\bar{\delta}}{a}.
\end{align}
Next, to obtain the critical imaginary vector potential $h_c$ and the change of the energy gap $\Delta(ih)$, we introduce the imaginary vector potential as $k\rightarrow k-ih$ and then the energy eigenvalues become 
\begin{widetext}
\begin{align}
    E_\pm(k, \delta, h)= \pm t \sqrt{2 (1+\bar{\delta}^2) + 2 (1 - \bar{\delta}^2) (\cosh 2ah \cos 2ak + i \sinh 2ah \sin 2ak)}.
\end{align} 
\end{widetext}
The critical value is obtained from the condition of $E_+(\pm \pi/(2a), \delta, h_c) - E_-(\pm \pi/(2a), \delta, h_c) = 0$ and then we obtain the critical value $h_c$ as
\begin{align}
h_c = \frac{1}{2a} \mathrm{arcosh}\left( \frac{1 + \bar{\delta}^2}{1-\bar{\delta}^2}\right).
\end{align}
$\Delta(ih)= E_+(\pm \pi/2, \delta, h) - E_-(\pm \pi/2, \delta, h)$ is obtained as
\begin{align}
\Delta(ih) = 4t \sqrt{ \frac{1+\bar{\delta}^2}{2} - \frac{1 - \bar{\delta}^2}{2} \cosh 2ah}.
\end{align}

As shown in Figs.~\ref{Fig:lattice}~(a-1) and (a-2), the above quantities show good agreement with Eqs.~(\ref{eq:h_c}) and (\ref{eq:Delta(ih)}) in a broad range from the weak coupling regime where $\bar{\delta}$ is small. Furthermore, it can be checked analytically within the weak coupling approximation ($\bar{\delta} \ll 1$). For Eq.~(\ref{eq:h_c}), using the relation
\begin{align}
\frac{1}{2} \mathrm{arcosh} \left(\frac{1+x^2}{1-x^2}\right) = x + \frac{x^3}{3} + \mathcal{O}(x^5),
\end{align}
we take terms up to the first order in $\bar{\delta}$ and then arrive at $h_c=\bar{\delta}/a=(e/e^*)\Delta_0/(2 v)$. This is nothing but Eq.~(\ref{eq:h_c}). As for Eq.~(\ref{eq:Delta(ih)}), focusing on the small-$h$ regime and using $\cosh 2ah \sim 1 + 2(ah)^2$, we can obtain $\Delta (ih) \sim 4t \sqrt{\bar{\delta}^2- (ah)^2 + \bar{\delta}^2 (ah)^2} \sim 4t \sqrt{\bar{\delta}^2- (ah)^2} = \Delta_0 \sqrt{1 - (ah/\bar{\delta})^2}\sim \Delta_0 \sqrt{1-(h/h_c)^2}$. Then, we analytically derive Eq.~(\ref{eq:Delta(ih)}) within the weak-coupling approximation.

\subsection{Spinless fermions with a nearest-neighbor interaction (XXZ spin chain)}

The model Hamiltonian is
\begin{align}
H_\mathrm{SLF}  &= - t \sum_{i} \left( c^\dagger_{i+1} c_i + \mathrm{h.c.} \right)\nonumber \\
&\qquad + V \sum_i
\left( c^\dagger_{i+1} c_{i+1} - \frac{1}{2} \right) \left( c^\dagger_{i} c_{i} - \frac{1}{2} \right). \label{eq:Ham_SLF}
\end{align}
This model describes fermions with a nearest-neighbor interaction. Here we consider the strong repulsive interaction $V>2t$ and the half-filled case. Under these conditions, the ground state shows charge-density wave (CDW) and becomes gapped. This model is mapped to an XXZ spin chain by means of Jordan-Wigner transformation $S^+_i = c^\dagger_i \exp \left( i \pi \sum_{j = -\infty}^{i-1} c^\dagger_j c_j  \right)$ and $S^z_i = c^\dagger_i c_i - 1/2$ with a canonical transformation $c_i \to (-1)^i c_i$ as
\begin{align}
H_\mathrm{XXZ}&= J \sum_i \left\{ S^x_{i+1} S^x_i + S^y_{i+1} S^y_i + \Delta S^z_{i+1} S^z_i \right\} \nonumber \\
&= J \sum_i \left\{ \frac{1}{2} (S^+_{i+1} S^-_i + S^-_{i+1} S^+_i) + \Delta S^z_{i+1} S^z_i \right\} \label{eq:Ham_XXZ}.
\end{align}
Here $J = 2 t$ and $\Delta = V / 2t$. The XXZ spin chain exhibits an antiferromagnetic order for $\Delta >1$, which corresponds to the CDW order of the original fermionic model (\ref{eq:Ham_SLF}). With bosonization techniques, the low-energy effective theory of this model is given as
\begin{align}
H_\mathrm{XXZ} &= \frac{v}{2 \pi} \int dx \left\{ K(\pi \Pi)^2 + \frac{1}{K} (\nabla \phi)^2 \right\}\nonumber\\
&\qquad+ g \int dx \cos (4 \phi (x)). 
\end{align}
Here, $v$, $K$, and $g$ take renormalized values due to the cosine term. We find that the parameter $\beta$ appearing in Eq.~\eqref{eq:sine_Gordon_S_with_h} is equal to 4 and thus the elementary excitation has a fractional charge $-e^*=-e/2$, which corresponds to a spinon excitation in the XXZ model.

To confirm Eqs.~(\ref{eq:h_c}) and (\ref{eq:Delta(ih)}), we calculate the energy gap $\Delta_0$,  the velocity $v$, the critical value $h_c$, and the change of the energy gap $\Delta(ih)$. This model is known to be exactly solvable with Bethe ansatz~\cite{Takahashi_book} and thus we can obtain the above quantities.

First, we consider the energy gap $\Delta_0$. The dielectric breakdown in the sine-Gordon model corresponds to the soliton-antisoliton pair creation and thus it corresponds to \textit{spinon-antispinon excitation}. The energy spectrum of the spinon excitation is given as
\begin{align}
\mathcal{E}(p) = \frac{J K(u) \sinh \gamma}{\pi} \sqrt{1-u^2 \sin^2 (ap)}, \label{eq:Excitation_XXZ}
\end{align}
where $\gamma = \mathrm{arcosh} \Delta$, $K(u)$ is the complete elliptic integral of the first kind and $u$ is its modulus defined as $K(\sqrt{1-u^2})/K(u)=\gamma/\pi$~\cite{Takahashi_book}. This spectrum takes the lowest value at $p=\pm \pi/(2a)$. Since the energy spectrum of the antispinon takes the same form, the energy gap of the spinon-antispinon excitation is 
\begin{align}
\Delta_0 = \frac{2 J K(u) \sinh \gamma}{\pi} \sqrt{1-u^2}. \label{eq:Delta0_XXZ}
\end{align}
Next, we derive the velocity $v$ which appears in the sine-Gordon theory (\ref{eq:sine_Gordon_S_with_h}). We can obtain the velocity $v$ from the above energy spectrum (\ref{eq:Excitation_XXZ}). We expand the energy spectrum around $p = \pi/(2a)$ where the excitation energy takes the minimum value. Then we obtain $\mathcal{E}(p) = (J  K(u) \sinh\gamma / \pi) \{ \sqrt{1- u^2} + [a^2 u^2/(2\sqrt{1-u^2})] (p - \pi/(2a))^2 \} + \mathcal{O}[(p-\pi/(2a))^4] $. Comparing this result with a relativistic energy spectrum $\mathcal{E}_0 (p, 0) = \sqrt{m^2 + v^2 (p-\pi/(2a))^2} = m + [v^2/2m] (p-\pi/(2a))^2 + \mathcal{O}[(p-\pi/(2a))^4]$, we obtain the mass and the velocity as $m = (J K(u) \sinh \gamma/\pi) \sqrt{1-u^2}$ and
\begin{align}
    v= \frac{J a K(u) u \sinh \gamma}{\pi}. \label{eq:v_XXZ}
\end{align} 
From this, we can confirm that the mass $m$ corresponds to a half of the energy gap $\Delta_0$ and the velocity approaches the value of the spinon velocity in a Heisenberg spin chain, $v_s = \pi J a /2$ when taking the limit of $\Delta \to 1$. 

As shown in previous studies~\cite{Albertini1996}, the model is still solvable with an imaginary vector potential and thus we can also calculate the critical value of the imaginary vector potential $h_c$ and the energy gap with the imaginary vector potential $\Delta(ih)$ analytically. Following Refs.~\cite{Albertini1996, Nakamura2006}, the critical value is given as
\begin{align}
    h_c = \frac{2}{a}~\mathrm{arcosh}\left(\frac{1}{u}\right), \label{eq:hc_XXZ}
\end{align}
and the energy gap is written as
\begin{align}
    \Delta(ih) = \frac{2 J  K(u) \sinh \gamma}{\pi} \sqrt{1-u^2 \cosh^2 \left(\frac{ah}{2}\right)}~. \label{eq:Delta(ih)_XXZ}
\end{align}

As shown in Figs.~\ref{Fig:lattice}~(b-1) and (b-2), these quantities $\Delta_0$, $v$, $h_c$, and $\Delta(ih)$ satisfy the relations (\ref{eq:h_c}) and (\ref{eq:Delta(ih)}) in a broad range including the weak coupling regime around the isotropic point $\Delta = 1$. Furthermore, Eq. (\ref{eq:h_c}) in the weak coupling regime can be analytically confirmed as follows. Using Eqs.~(\ref{eq:Delta0_XXZ}) and $\beta = 4$, (\ref{eq:v_XXZ}) and (\ref{eq:hc_XXZ}), it is shown that $[(e/e^*)\Delta_0/(2v)]/h_c = [\Delta_0/v]/h_c = [\sqrt{1-u^2}/u]/\mathrm{arcosh}(1/u) \to 1$ with $u \to 0$. 

\subsection{Fermi-Hubbard chain}

The one-dimensional fermionic Hubbard model is written as
\begin{align}
H_\mathrm{Hub}=- t \sum_{i\sigma} ( c^\dagger_{i \sigma} c_{i+i \sigma} + \mathrm{h.c.}) + U \sum_i n_{i \uparrow} n_{i \downarrow}.
\end{align}
The ground state at half-filling is a Mott insulator for $U>0$. Let us apply bosonization techniques to this model. This is a model of fermions with spins and thus the effective theory has the charge sector and the spin sector. Since we are interested in the breakdown phenomena due to the electric field, we focus only on the charge sector and omit the subscript for the charge sector $\rho$ below. The Hamiltonian of the charge sector is written as 
\begin{align}
H_\mathrm{Hub} &= \frac{v}{2 \pi} \int dx \left\{ K(\pi \Pi)^2 + \frac{1}{K} (\nabla \phi)^2 \right\} \nonumber \\
&\qquad + g \int dx \cos (2\sqrt{2} \phi (x)). 
\end{align}
Here, $v$, $K$, and $g$ take renormalized values due to the cosine term. The cosine term represents the umklapp process and makes the system gapped. The parameter $\beta$ is $2\sqrt{2}$ for this model. Since $\phi$ is a field for the charge sector and thus the definition of $e^*$ is modified as $e^*=2\sqrt{2}e/\beta$, the elementary excitation corresponding to the soliton in the sine-Gordon model has a charge $-e^* = -e$ which represents a doublon excitation (a double occupancy) in the Hubbard model. 

To confirm Eqs.~(\ref{eq:h_c}) and (\ref{eq:Delta(ih)}), we calculate the energy gap $\Delta_0$, the velocity $v$, the critical value $h_c$, and the change of the energy gap $\Delta(ih)$. This model is also known to be exactly solvable with Bethe ansatz~\cite{Lieb1968, Essler_book} and thus we can obtain the above quantities.

The energy (charge) gap is calculated using Bethe ansatz as
\begin{align}
\frac{\Delta_0}{4t} = u - \left[ 1 -  2 \int^\infty_0 d \omega \frac{J_1(\omega)}{\omega (1 + e^{2u|\omega|})} \right],
\end{align}
with $u= U/4t$ where $J_n (x)$ is the $n$-th Bessel function~\cite{Lieb1968}. 
The velocity of the charge mode $v$ is also calculated based on the Bethe ansatz in the same way as in the XXZ model as discussed in the previous subsection~\cite{Essler_book}. The result is
\begin{align}
\frac{v}{4ta} = \frac{
\left[
-1+u+2\int^\infty_0 \frac{J_1 (\omega)}{\omega (1+e^{2u \omega})}
\right]^{\frac{1}{2}}
\left[
1-2\int^\infty_0 \frac{\omega J_1 (\omega)}{1+e^{2u \omega}}
\right]^{\frac{1}{2}}
}{
2 \left[
1-2\int^\infty_0 \frac{J_0 (\omega)}{1+e^{2u \omega}}\right]
}.
\end{align}

As shown in a previous study~\cite{Fukui1998}, the model is still solvable with an imaginary vector potential and thus we can also calculate the critical value of the imaginary vector potential $h_c$ and the energy gap with the imaginary vector potential $\Delta(ih)$ analytically. Following Refs.~\cite{Fukui1998, Nakamura2006, Oka2010, phdthesis_Nakamura}, the critical value is
\begin{align}
h_c &= \frac{1}{a} \left[ b_c - i \int^{\infty}_{-\infty} d \lambda \theta (\lambda + i \sinh b_c) \sigma_0 (\lambda) \right], 
\end{align}
where $b_c=\mathrm{arsinh} (u)$, $\theta(x) = -2 \arctan \left(x/u\right) $, and $\sigma_0 (\lambda) = (1/2\pi) \int^\infty_0 d\omega \left[ J_0(\omega) \cos(\omega \lambda) /  \cosh (2u \omega) \right]$.
The change of the energy gap $\Delta(ih)$ is
\begin{align}
\frac{\Delta(ih)}{4t} &= u - \cosh(a\kappa(h)) \nonumber \\
& \qquad + 2 \int^{\infty}_{0} d\omega \frac{\cosh(\omega \sinh (a\kappa(h))) J_1 (\omega)}{\omega (1+e^{2u\omega})}, 
\end{align}
where $\kappa (h)$ is defined from the following relation:
\begin{align}
    h(\kappa) = \kappa - \frac{2}{a} \int^\infty_0 d\omega \frac{\sinh(\omega \sinh(a \kappa))J_0(\omega)}{\omega(1+e^{2u \omega})}
\end{align}
As shown in Figs.~\ref{Fig:lattice}~(c-1) and (c-2), these quantities $\Delta_0$, $v$, $h_c$, and $\Delta(ih)$ satisfy the relations (\ref{eq:h_c}) and (\ref{eq:Delta(ih)}) in a broad range including the weak coupling regime (i.e., small $U/t$).

\bibliographystyle{apsrev4-1}
\bibliography{ref.bib}

\begin{thebibliography}{75}
\bibitem [{FNA()}]{FNA}%
  \BibitemOpen
  \href@noop {} {}\bibinfo {note} {We neglect the degeneracy due to the
  internal degrees of freedom that do not couple to the electric field, such as
  spins, and concentrate on the charge excitations}\BibitemShut {NoStop}%
\bibitem [{FNE()}]{FNE}%
  \BibitemOpen
  \href@noop {} {}\bibinfo {note} {In the action, the integral over the
  imaginary time $\tau$ is performed from $0$ to $\infty$. Since we consider a
  sufficiently large system, the integration range along the $x$-direction is
  taken from $0$ to $\infty$ in the thermodynamic limit.}\BibitemShut {Stop}%
\bibitem [{FN5()}]{FN5}%
  \BibitemOpen
  \href@noop {} {}\bibinfo {note} {For spinless systems, the physical charge of
  solitons is given by $Q_\mathrm{phys}=-(1/\pi)\int dx \partial_x \phi$
  (e.g.~\cite{Giamarchi_book}) and thus this is related to the topological
  charge as $Q_\mathrm{phys}=- 2/\beta Q_\mathrm{top}$. Therefore, the physical
  charge of a soliton is $- e^* = - 2e/\beta$. Note that $e^* =
  2\sqrt{2}e/\beta$ when we use $\beta$ of the charge sector Hamiltonian of
  spinful systems. It is because the physical charge is defined as
  $Q_\mathrm{phys}=-(\sqrt{2}/\pi) \int dx \partial_x \phi_\rho$ where
  $\phi_\rho$ is a field of the charge sector. For the definition of
  $\phi_\rho$, see Appendix~\ref{sec:bosonization}}\BibitemShut {NoStop}%
\bibitem [{FNF()}]{FNF}%
  \BibitemOpen
  \href@noop {} {}\bibinfo {note} {When $\phi$ is a field for the charge sector
  in a spinful model (e.g. Fermi-Hubbard model), the formula~(\ref{eq:h_c})
  still holds, but the definition of $e^*$ is modified as $e^* =
  2\sqrt{2}e/\beta$~\cite{FN5} because the vector potential is introduced as
  $\sqrt{2}h/\pi \int d\tau dx (\partial_\tau \phi)$ in
  Eq.~(\ref{eq:sine_Gordon_S_with_h}) in the case of the charge
  sector}\BibitemShut {NoStop}%
\bibitem [{FNJ()}]{FNJ}%
  \BibitemOpen
  \href@noop {} {}\bibinfo {note} {The non-Hermitian lattice Hamiltonian
  [Eq.~(\ref{eq:lattice_model}) with $e^{\pm iaA} \rightarrow e^{\mp ah}$] is a
  real matrix and thus its eigenvalues appear as complex-conjugate pairs.
  Therefore, the partition function $Z$ is real}\BibitemShut {NoStop}%
\bibitem [{FNC()}]{FNC}%
  \BibitemOpen
  \href@noop {} {}\bibinfo {note} {While we only show the examples of $e^*/e =
  1/2$ or $1$ in this study, we can construct the case of $e^*/e = 1/n$ using
  other types of CDW insulators, where one electron sits per $n$ sites ($n$:
  positive integer)}\BibitemShut {NoStop}%
\bibitem [{FNG()}]{FNG}%
  \BibitemOpen
  \href@noop {} {}\bibinfo {note} {Note that the signs of the deviation from
  the field-theoretical prediction in Fig. 3 (b-1) and (c-1) are different from
  that of Fig. 3 (a-1). It strongly depends on the details of the lattice
  model, and we leave it as an open question.}\BibitemShut {Stop}%
\end{thebibliography}
\end{document}